# Crossed chiral band approximation for wide-band self-collimation of light

Melike Gumus[1*], Onder Akcaalan[2], and Hamza Kurt[1]

[1]Department of Electrical and Electronics Engineering, TOBB University of Economics and Technology, Ankara 06560, Turkey
[2]Department of Physics, Bilkent University, Ankara 06800, Turkey
*E-mail: gms.mlke@gmail.com

**Abstract**

We propose a perspective to the evaluation of the wide bandwidth phenomenon, by introducing the band tailoring and chiral band approximation on the self-collimation effect for low-symmetric photonic structures. In the case of the crossing of the bands, we claim the excitation of the lower mode can provide the utilization of the entire bandwidth by suppressing the intersection regions where the frequencies tend to mix. Thereby, we design broadband self-collimation capable, defect-free photonic structures and examine their performances. A fractional bandwidth of 0.35 ($a/\lambda = 0.429 - 0.607$) and 0.37 ($a/\lambda = 0.481 - 0.701$) are achieved for radii of $r = 0.25a$ and $r = 0.23a$, respectively. We explore the full-range collimation using transmission and E-field intensity analysis in addition to band diagrams and group velocity dispersions. Moreover, we indicate all-angle collimation validity even for highly tilted sources up to an angle of $80°$.

## 1. Introduction

Photonic crystals (PhCs) are structures that provide light propagation control by refractive index modulation [1,2]. Since these structures are able to work in a wide range of wavelength spectrums due to their scalability, they have been widely used in optical and photonic researches. Scalability features allow a design to be easily used for different wavelengths. During PhC investigations, some peculiar light properties, such as super-prism [3], high-quality factor cavity [4], and self-collimation [5] have been discovered. When the investigations are considered in terms of applications, it can be mentioned that each phenomenon can be utilized for a device such as wavelength-division multiplexers (super-prism), sensors (cavity), photonic integrated circuits (self-collimation). The usage of these properties brings along some common operational constraints and figure of merits. Bandwidth (BW) is one of these figure of merits due to information processing capability about the working interval of a phenomenon for photonic applications and the wider BW means more operating frequency to process data. As a simple mathematical description, the division of the interval between ending and beginning frequencies to the center frequency describes the fractional bandwidth (FBW). In this point of view, the FBW can be determined through the band diagram branches and equifrequency contours (EFC's) which give direct information of the light propagation in the frequency domain. The utilization importance of BW with self-collimation ability can be exemplified by investigations on hybrid plasmonic PhCs [6], flat band localization [7], and diffraction management [8].

## 2. Numerical investigations of proposed structures

*2.1. Proposed designs and WBW focused frequency domain analysis*

In this study, the self-collimation phenomenon, which is widely used for various photonic applications [9–11], is chosen to investigate the BW characteristics, because it provides a comprehensive analysis possibility by employing EFC and band diagram. Considering EFCs, flat contours give insight about the collimated light propagation of a structure. Due to Poynting vector (group velocity) conservation rules, a vector which comes across to the normal of a flat contour goes on its way without any deflection [12]. It means the light deflection along the latarel direction will be prevented if it comes through the normal direction with the flat contour frequency [13,14]. In the condition of the complete flat EFCs, they have a linear dense alteration of frequency along the $\Gamma - X$ direction, besides all frequencies varying along this direction, must be constant along the $X - M$ direction. It means the light has a self-collimation tendency for all incident angles from zero to $90^0$. Meanwhile, it is expected that $M - \Gamma$ has the same attitude (linear alteration) with $\Gamma - X$ in terms of frequency. Considering the band, the mentioned flatness corresponds to a branch-flat-branch (B-F-B) manner throughout $\Gamma - X - M - \Gamma$, respectively. The dense EFCs come to exists as high-slope branches in band diagrams and corresponds to the wider BW with respect to other band structures. Furthermore, the mentioned B-F-B tendency reveals to a chiral band relation between two adjacent bands, which superimpose along the $X - M$ direction, thereby it enables a significant BW addition to the all-angle property. This combined and well-performed band structure has been named as the *chiral* band in the study, see Fig. 1(c). To achieve that kind of fully adequate band, there is a need to control the band branch behavior. Low-symmetric structures allow more controllable mechanism which provides more degree of freedoms. Therefore, we choose to direct the investigations towards C2 symmetric configuration because of its simplicity [15]. The mentioned C2 symmetric configuration enables the dielectric density to be distributed by altering its location or radius, and diversity of dielectric density leads to control of branch behavior and thereby the operating frequency interval. At this point, it should be remarked that each of the frequency domain analyses, which is to be focused on, are performed in MATLAB [16] by solving the eigenvalue problems by way of plane wave expansion (PWE) method as mentioned in our previous studies [15,17]. The first step to band investigations is the alteration of the rotation angle $\theta_{rot} = 0°, 30°, 60°, 90°$ of the elements in the unit cell for the 3rd TM band along the $\Gamma - X$ direction in the vicinity of $\Gamma$ symmetry point. In this condition, taking the filling ratio (0.3927) constant for each unit cell, we alter the rotation angle by reference to the center of the cell. Then the configurations are compared and found a convenient unit cell with highest FBW value for promising self-collimation. Due to the realization of an inverse proportional relation between the rotation angle of the rods and slope of the branches, one can say the rotation angle $\theta_{rot} = 0°$ corresponds to the highest FBW value, see the upper part of Fig 1(a). The second step is to change the element radii to obtain the highest FBW value once again. In this step, while the rotation angle is $\theta_{rot} = 0°$, the radii are altered as $r = 0.05a_{r_1}, 0.10a_{r_2}, 0.15a_{r_3}, 0.20a_{r_4}, 0.25a_{r_5}$, where $a$ is a lattice constant, without changing the center points of elements, seen in the lower part of Fig. 1(a). Increasing filling factor causes dispersion branch to stay along at a lower frequency interval. Thereby, it is properly proved that the band tailoring is possible by changing the filling ratio in the unit cell. When the altered radii are compared, it is observable that the high FBW value is obtained for $r_5 = 0.25a$ radii rods which are positioned to each other with $\theta_{rot} = 0°$ rotation angle, thereby, the optimum unit cell is determined. The structure which is designed by utilizing the mentioned unit cell is shown in Fig. 1(b), it has the extends for $x$ and $y$-axes $x = 40a, y = 10a$, respectively. The elements in the unit cell are rod type alumina dielectrics with $\varepsilon = 9.8$ in a square-lattice, an extended depiction of the unit cell is shown in Fig. 1(b) as an inset. To achieve the *chiral* band structure mentioned above, additional fine tuning is performed in terms of radii, and finally, a perfect *chiral* band is obtained for $r = 0.23a$. Since reducing the radii of the elements in the unit cell causes a smaller filling ratio with respect to $r = 0.25a$, the bands are observed at higher frequencies. A comparison between two different radii can be seen in Fig. 1(c). According to Fig. 1(c), $r = 0.23a$ and $r = 0.25a$ radii configurations behave as *chiral* (symmetric) and *chiral-like* (nearly symmetric) bands, respectively. Also, in Fig. 1(c) some remarkable representations can be seen such as wide bandwidth region (WBW) and superimpose region. The mentioned regions identify the mirror-symmetric positioning and superposition of the adjacent bands, respectively. Adjacent bands are symmetric with respect to mirror lines which consist of superimpose region and divide the WBW region into equal (*chiral*) or nearly equal parts (*chiral-like*) in terms of frequency. The branches extend across $\Gamma - X$ and provide ~0.36 FBW which is a satisfactory result. Considering the other bands besides *chiral* ones, an extensive remark can be attained for band behaviors. The 2D and 3D versions of TM band diagrams of $r = 0.25a$ and $r = 0.23a$ are illustrated in Figs. 1(d) and 1(e), respectively. Here, the 3D presentation focuses on the bands from the 3rd to 7th along the $\Gamma - X$, which is the investigation direction of self-collimation. When two different radii configurations are compared, it is seen that the bands show similar characteristics, but the drastic change occurs for bands of (3rd-6th) / (3rd-7th) for $(r = 0.25a)/(r = 0.23a)$ with respect to the $\Gamma$ symmetry point. For the configurations of $r = 0.23a$ and $r = 0.25a$, there are three and two crossing points, respectively. These crossing points cause intersected frequency regions which include the same frequencies on the bands. As a consequence of crossing, some light propagation disturbances may occur in the intersection regions due to mixed modes.

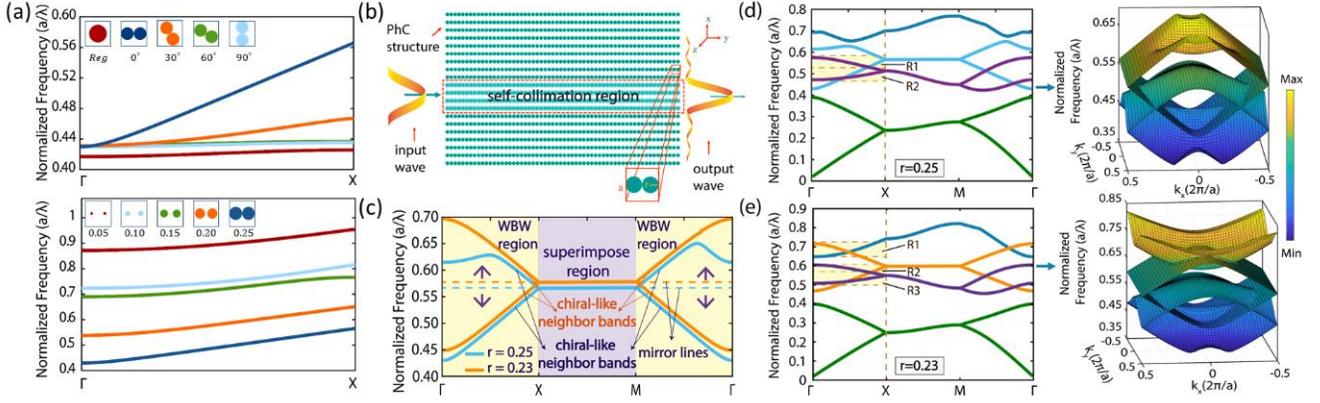

**Figure 1.** Calculated band diagrams for ideal structure investigation regarding rotational and radii manipulations at up and down figures of (a), respectively. (b) Schematic of the self-collimation region with unit cell projection representation. (c) *Chiral-like/Chiral* band representation of the 3rd-6th/3rd-7th bands. (d) and (e) are the TM band diagrams of the ideal structure along $\Gamma - X - M - \Gamma$ direction for $r = 0.25a$ and $r = 0.23a$, respectively. Shaded parts show the intersection regions. $\Gamma - X$ oriented 3D band representations are positioned nearby each diagram.

Another indicator of self-collimation property is GVD analysis as well. The relation between GVD, dispersion diagram and self-collimation can be explained as follows; a linear dispersion branch has a constant slope which states the $k$-vector components are not affected by altering frequencies. This relation prevents the different speed of $k$-vector components and de-phasing, thus, results in self-collimation [18,19]. As a simple mathematical definition, the GVD's of the structures are calculated through the relation of $GVD =$

$\partial/\partial w\,(1/v_g) = \partial^2 k/\partial w^2$. The equation indicates, GVD is the frequency-dependent group velocity alteration, and this dependency induces beam broadening which caused by phase shifts on wave vector. The fact the light does not broaden over time is an indicative of self-collimation. Thus, GVD values are expected to be at the vicinity of zero. GVD values in a logarithmic scale, annihilate the demonstration limitation and enable a wider scale investigation of the full frequency range, see Fig. 2(a) and Fig. 2(b). Even if GVD values are reasonable for a big portion of the BW, by using logarithmic scale, values are determined quite large contrary to expectation at the beginning frequencies of intersection region 1 (R1) $GVD_{r=0.25a} = -10^8 / GVD_{r=0.23a} = -10^8$ and ending frequencies of intersection region 2 (R2) $GVD_{r=0.25a} = -10^7 / GVD_{r=0.23a} = -10^7$. Additionally, an intersection region represented as R3 emerges for $r = 0.23a$ structure. This region, has a large GVD value of $GVD_{r=0.23a} = -10^7$ at the beginning frequencies, as in the other regions. The emergence of a new intersection region provides an opportunity for the structure comparison.

Despite the common idea, further investigation should be performed because to decide the propagation of light by only applying to the GVD is a limited perspective. GVD is the second derivative of the band diagram hence, there is a direct relation between GVD and band diagram. Therefore, a missing point in the band diagram will be valid for the GVD as well. In such cases, time-domain analysis can be used as supportive of the frequency domain analysis. Sometimes, large GVD values can be deceptive, it should be remarked that even if the GVD may diverge to higher values, the structure may keep the collimation ability practically. Our study emphasizes the fact that by the Gaussian source excitation of the low order mode, which supports the collimation, the effects of the other crossing modes can be suppressed [20]. Thereby, encountered mixing of modes problems can be eliminated and the entire BW can be exploited. In Figs. 2(a) and 2(b) each line demonstrates calculated GVD values of the dispersion bands. Shaded parts on bands correspond to frequency intersection regions, and the green dashed lines are the limits of the region that the self-collimation phenomenon is valid.

### 2.2. WBW focused time domain analysis

The valid collimation region should be determined according to the field intensity profiles besides transmission analysis with adequate data points. Although GVD is quite high for crossing points, high transmission appears on the entire frequency range without any interruption, according to Fig. 2(c), hence our claim is supported. In Fig. 2(c), the transmission results do not indicate any drop points which means dispersion bands are definitely crossing without any gap. Transmission implies good self-collimation ability for the entire frequency interval which is illustrated by green and yellow shaded rectangles for $r = 0.25a$ and $r = 0.23a$, respectively. For both configurations, the average transmission is 70% and starts to decline at the frequencies which are approaching to the edges of the operation frequency interval. This information supports our *chiral* band approximation by providing ~0.36 wide FBW, and strengthens the low order mode excitation approach which is not influenced by the intersection distortion.

In order to strengthen the approach shown by transmission, the light propagation in the structure in terms of ***E***-field intensity profiles is examined by utilizing finite-difference-time-domain (FDTD) method with LUMERICAL software [21]. Sampling frequencies are $a_{0.25}/\lambda = 0.429, 0434, 0.500, 0.595, 0.607$ and $a_{0.23}/\lambda = 0.481, 0.491, 0.601, 0.691, 0.701$, which refer to the self-collimation stages, chosen. Additionally, chosen frequencies to clarify that the modes do not mix in the intersection region are $a_{0.25}/\lambda = 0.472, 0575$ and $a_{0.23}/\lambda = 0.508, 0.604, 0.647$.

Results indicate the self-collimation effect is sustainable even for intersection regions throughout the proposed interval, see the results in Fig. 3(a,b). While sampling frequencies are determined from the beginning, center, and ending of the self-collimation region to observe propagation tendency, the adjacent frequencies are selected to represent the light behavior near the limits of the region. The most problematical regions have been labeled as intersection regions which are shown the beginning frequencies of R1 and R3 regions and ending frequencies of R2 region in the investigation of GVD. The intersection frequencies are chosen as the near points of these regions. By means of the selected frequencies, the coupling and the strengthening process of the light are shown step by step in the intensity profile representations. If the intensity profiles associate with the bands in Fig. 1(e), it is seen that stronger collimation emerges for the frequencies at the linear part of the branch. When the observations are made by considering the frequencies, it can be said that more satisfying collimation is obtained for the adjacent frequencies which are relatively closer to the center of branch with respect to beginning and ending frequencies of the self-collimation branch for configurations. Here, adjacent frequencies are $a_{0.25}/\lambda = 0.434, 0.595$ and $a_{0.23}/\lambda = 0.491, 0.691$ while beginning and ending frequencies are $a_{0.25}/\lambda = 0.429$ / $a_{0.23}/\lambda = 0.481$ and $a_{0.25}/\lambda = 0.607$, $a_{0.23}/\lambda = 0.701$. First, the light reaches the self-collimation region at the beginning frequencies then it demonstrates a reasonable self-collimation at adjacent frequencies. After that, light reaches the strongest self-collimation in the vicinity of center frequency $a_{0.25}/\lambda = 0.500$, $a_{0.23}/\lambda = 0.601$. Finally, it completes its propagation at the band edge by reaching the ending frequencies. At $a_{0.23}/\lambda = 0.691$ and $a_{0.23}/\lambda = 0.701$ frequencies, it undergoes localizations due to the slow-light effect. Thereby, a rare self-collimated slow-light is observed for these frequencies which correspond to more flat parts of the band diagram. The results in Fig 3(a,b) prove our claim by supporting a self-collimator with the widest BW range, including intersection frequencies such as $a_{0.25}/\lambda = 0.472, 0.575$ and $a_{0.23}/\lambda = 0.508, 0.604, 0.647$.

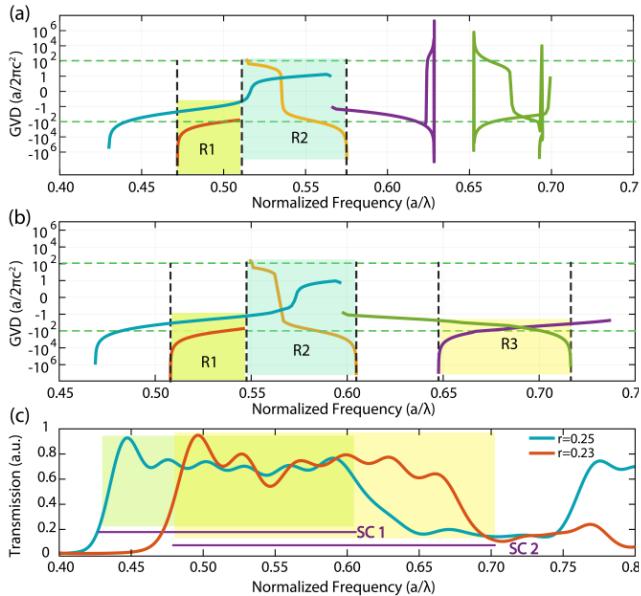

**Figure 2.** (a) and (b) are calculated GVD characteristics for direction for $r = 0.25a$ and $r = 0.23a$, respectively. (c) The full-range transmission spectrum of the structure, shaded SC1 and SC2 parts indicate self-collimation regions of $r = 0.25a$ and $r = 0.23a$, respectively.

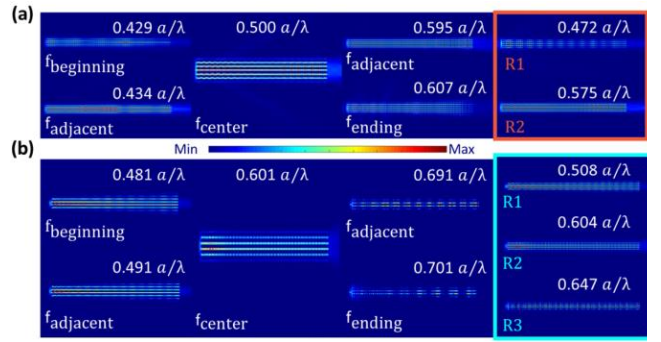

**Figure 3.** *E*-field intensity profiles at the selected normalized frequencies with intersection region representations for (a) $r = 0.25a$ and (b) $r = 0.23a$.

As an additional feature, the all-angle performance, which upgrades the collimation to super-collimation, can be used as supportive for device application idea [14]. In the study, the field intensities are performed for $\theta_{source} = 0°, 20°, 40°, 60°, 80°$ source, and results verify the self-collimation, see Figs. 4(a)-4(e), respectively. According to the achieved results, the self-collimation behavior is valid even at huge angles for both structures. Here, $a_{0.25}/\lambda = 0.500$ and $a_{0.23}/\lambda = 0.601$, each of which is the center frequencies for their frequency intervals, are chosen as the operating frequencies for all-angle self-collimation investigations. The results are quite similar for both structures at each tilted angle illustration, and inset representations indicate results of $r = 0.23a$. Despite the fact that coupling performance decreases proportionally with increasing tilt angles, our structure provides a good coupling and maintains collimation ability. However, the coupling performance of light to the structure can be improved by using anti-reflection elements such as smaller dielectric rods [22].

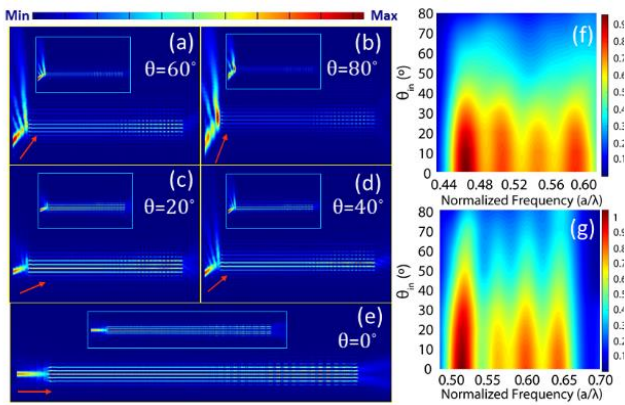

**Figure 4.** *E*-field intensity profiles for tilted source at $a_{0.25}/\lambda = 0.500$ and $a_{0.23}/\lambda = 0.601$(inset) normalized frequencies. (a), (b), (c), (d), and (e) illustrate sources at $\theta_{source} = 0°, 20°, 40°, 60°, 80°$ tilt angles, respectively. Transmission maps of the structures in terms of the incident angles $\theta_{source} = 0°, 20°, 40°, 60°, 80°$ and normalized frequencies which indicate the intervals of their self-collimation regions for the cases of (f) $r = 0.25a$ and (g) $r = 0.23a$.

When the transmissions of incident angles are investigated, it can be claimed that some coupling issues seem more apparent. The transmission results for all-angle is given in Fig. 4(f, g). The transmission maps indicate their maximum values of $T_{0.25} = 92\%$ and $T_{0.23} = 95\%$ for $r = 0.25$ and $r = 0.23$, respectively at $\theta_{source} = 0°$. These values decrease to $T_{0.25} = 26\%$ and $T_{0.23} = 12\%$ for a minimum transmission. Additionally, the minimum values emerge for both structures at $\theta_{source} = 80°$ as expected due to the strong back-reflections at the interface of structures. For each frequency value, all transmission values ranging from $0°$ to $80°$ fall to approximately their 1/3rd value. Thereby, values are obtained as approximately maximum $T_{0.25} = 35\%$ and minimum $T_{0.23} = 3\%$ for $\theta_{source} = 80°$.

### 3. Conclusion

As a conclusion, a WBW approach has been emphasized in this study by exploiting the self-collimation phenomenon. The first optimization of the rotation and the radii of the high-symmetric unit cell elements has been provided by band tailoring along $\Gamma - X$ direction. Thus, a more convenient band tendency has been obtained. Then, two convenient configurations which show *chiral-like* and *chiral* behavior, have been selected due to different chirality characteristics. According to the GVD results, it has been estimated some common frequencies that emerged by the band crossing, may be problematic for the continuity of the self-collimation properly. Nevertheless, when FDTD analyses are conducted for the transmission and the field intensity profiles of both structure, it has been observed the self-collimation has not been destroyed for intersection regions. After that, by determining the self-collimation according to the *E*-field profiles the collimation interval has been identified as $a_{0.25}/\lambda = 0.429 - 0.607$ and $a_{0.23}/\lambda = 0.481 - 0.701$. According to the reasonable transmission results which fluctuate in the vicinity of 70% for both configurations, the self-collimation region has ~0.36 FBW value and the results have been supported by *E*-field intensity profiles. To show all-angle ability, using a source with the following tilted angles, $\theta_{source} = 0°, 20°, 40°, 60°, 80°$ have been chosen and the results indicated the validity of the self-collimation. It means that the designed structures have satisfactory transmission values for the annihilation of the alignment problem since the collimation is independent of the incidence angle. Analysis proves that the designed structures provide the widest bandwidth self-collimation by protecting its high transmission in a big portion of the full operation range due to its *chiral* property in spite of the band-crossing for optical interconnects, bends/splitters, optical-switches, and logic gates.

### Acknowledgements

The authors M. Gumus and H. Kurt acknowledge funding of the Scientific and Technological Research Council of Turkey (TUBITAK) with Project No. 115R036. H. Kurt also acknowledges the partial support of the Turkish Academy of Sciences (TUBA).